\documentclass[aip,reprint,floatfix]{revtex4-2}
\usepackage{graphicx} 
\usepackage{amsmath,amssymb,bm} 
\usepackage{siunitx}
\usepackage[hidelinks,colorlinks=true]{hyperref}

\begin{document}

\title{Demonstration of $0-\pi$ transition in Josephson junctions containing unbalanced synthetic antiferromagnets}
\author{D. Korucu}
\author{Reza Loloee}
\author{Norman O. Birge}%
\email{birge@msu.edu}
\affiliation{Department of Physics and Astronomy, Michigan State University, East Lansing, MI 48824, USA}
\date{\today}

\begin{abstract}
Josephson junctions containing ferromagnetic (F) materials have been the subject of intense study over the past two decades. The ground state of such junctions oscillates between 0 and $\pi$ as the thickness of the ferromagnetic layer increases. For some applications, it might be beneficial to replace a very thin F layer with an unbalanced synthetic antiferromagnet (SAF) consisting of two F layers of different thicknesses whose magnetizations are coupled antiparallel to each other. According to theory, such a system should behave similarly to a single F layer whose thickness is equal to the difference of the two F-layer thicknesses in the SAF. We test that theoretical prediction with Josephson junctions containing unbalanced Ni/Ru/Ni SAFs, keeping the thickness of one layer fixed at 2.0 nm and varying the thickness of the other layer between 2.0 and 5.0 nm.  We observe the first $0-\pi$ transition at a thickness difference of 0.86 nm, which closely matches the position of the transition observed previously using single Ni layers.  
\end{abstract}

\maketitle

Superconducting/ferromagnetic (S/F) hybrid systems have been the subject of intense study for the past two decades.\cite{buzdin_2005} S/F systems are interesting both because they exhibit a variety of new physical phenomena and because some S/F devices show promise for technological applications, e.g. in superconducting digital logic and memory.\cite{ryazanov_2012,soloviev_2017} The S and F materials in S/F hybrid systems interact in two distinct ways; we focus here on the novel proximity effects driven by the exchange field in F; we are not concerned with the so-called ``orbital" effects that result from the magnetic field produced by F. The primary physical mechanism underlying the proximity effects is the fact that the Cooper pairs in conventional S materials are in a singlet spin state.  When such a pair enters an adjacent F layer, the two electrons must enter different spin bands, which leads to a spatial oscillation of the pair correlation function.\cite{demler_1997}  In S/F/S Josephson junctions, those spatial oscillations may result in $\pi$-junctions, which have a ground-state phase difference of $\pi$ between the superconducting condensates of the two S electrodes.

This paper focuses on S/F${_1}$/N/F$_2$/S Josephson junctions, where F${_1}$ and F${_2}$ are ferromagnetic layers exchange-coupled antiferromagnetically by the intervening normal metal (N) layer.  The entire interior structure of the junction is called a synthetic antiferromagnet, or SAF. There have been several theoretical papers that deal with Josephson junctions with this general structure.\cite{bergeret_2001,bergeret_2001a,krivoruchko_2001,golubov_2002,barash_2002,chtchelkatchev_2002,zaitsev_2003,blanter_2004,vedyayev_2005,pajovi_2006,crouzy_2007,sperstad_2008,zaitsev_2009,robinson_2010} In junctions that contain no insulating layers, a central prediction of the theory is that the phase accumulated by a pair of electrons transiting through the first F layer, F${_1}$, is partially or fully cancelled when the same pair transits through F${_2}$, due to the fact that the majority and minority spin bands have swapped roles.\cite{blanter_2004}  That prediction is easy to understand in the limit of ballistic transport, where every electron travels straight through the F${_1}$/N/F$_2$ SAF without changing direction.  In the simplest case of a balanced SAF where F${_1}$ and F$_2$ are identical materials with the same thickness $d_F$, the phase cancellation results in the prediction that the critical current of such a junction is equal to that of a junction containing no ferromagnetic layers whatsoever!  That prediction is striking because the critical current in a ballistic S/F/S junction not only oscillates as a function of $d_F$, but also decays algebraically due to averaging over directions of trajectories.\cite{demler_1997,buzdin_1982} The phase cancellation removes both the oscillations and the algebraic decay.  Perfect phase cancellation is unrealistic in real systems, of course, due to scattering and unavoidable Fermi surface mismatches at the S/F and F/N interfaces.  Nevertheless, the prediction highlights the important role of phase cancellation in such a system. 

The situation in the diffusive limit is more subtle.  The critical current in a diffusive S/F/S junction also oscillates, but decays exponentially as a function of $d_F$.\cite{buzdin_1991}  In a diffusive junction containing a balanced SAF, the phase cancellation discussed above causes the oscillations to disappear, but the exponential decay remains.\cite{blanter_2004,crouzy_2007}  Hence the critical current of a diffusive S/F/S junction containing a balanced SAF is expected to decay exponentially as the total thickness of the SAF increases.  That is indeed what was observed in a recent study of junctions containing balanced Ni/Ru/Ni SAFs.\cite{mishra_2021}  With unbalanced SAFs, one expects to see oscillations as a function of the difference of the two F-layer thicknesses, along with an exponential decay depending on the sum of the thicknesses.\cite{blanter_2004} 

Surprisingly, the situation in unbalanced SAFs has not been fully studied experimentally.  The qualitative correctness of the physical picture outlined above has been confirmed in numerous studies of S/F${_1}$/N/F$_2$/S ``spin-valve" junctions containing two different F materials, where the relative orientations of the magnetizations can be toggled between parallel and antiparallel.\cite{bell_2004,robinson_2010,baek_2014,abdelqader_2014,baek_2015,gingrich_2016,baek_2017,niedzielski_2018,madden_2019,satchell_2020}  But obtaining a consistent quantitative fit to the data has proven to be difficult when F${_1}$ and F$_2$ are different materials.\cite{baek_2017,niedzielski_2018} In the present work both F layers are chosen to be Ni to enable us to make a quantitative test of the theory, which is the primary motivation for this work.  \textit{A priori}, we cannot predict if supercurrent transport in our junctions will fall into the ballistic or diffusive limit; earlier work on Ni junctions showed ballistic transport up to Ni thicknesses of several nm,\cite{robinson_2006,baek_2017,baek_2018} but scattering at the two Ni/Ru interfaces might cause our junctions to deviate from ballistic behavior. 

A secondary motivation for this study arose from our previous work on spin-valve junctions\cite{gingrich_2016,niedzielski_2018,madden_2019} as well as spin-triplet junctions,\cite{khasawneh_2011,klose_2012} which contain three F layers. We often use Ni as the ``fixed" magnetic layer in such junctions, because very thin Ni films have high coercive fields at low temperature and because Ni appears to be the best magnetic material for the transport of supercurrent.\cite{birge_2024}  The optimum Ni thickness for those applications is between 1.0 and 1.5nm.  Such thin Ni films have two major drawbacks.  First, the coercive field is so high (and grows as film thickness decreases) that initializing the magnetic state into a single-domain state requires applying a magnetic field of several hundred mT.\cite{klose_2012,baek_2014}  Second, there is indirect evidence that the remanent state of thin Ni nanomagnets with lateral dimensions of about 1 $\mu$m -- the typical size of Josephson junctions used in digital logic applications -- is not single-domain even after the high-field initialization. Our own experience with Ni suggests that increasing the thickness to just a few nm has a substantial positive impact on its magnetic behavior.  That suggests that replacing a very thin Ni layer with an unbalanced Ni SAF might improve its magnetic behavior, and still produce the Cooper pair phase shift needed to serve its purpose in spin-valve and spin-triplet junctions.  

To characterize the magnetic behavior of the Ni/Ru/Ni unbalanced SAFs, we deposited Nb(20)/Cu(2)/Ni($d_{F1}$)/Ru(0.9)/Ni($d_{F2}$)/Cu(2)/ multilayers (thicknesses are in nm) by dc triode sputtering on Si substrates in a vacuum system with base pressure $4 \times 10^{-6}$Pa.  The Ru thickness of 0.9 nm has been shown to maximize the antiferromagnetic coupling between the two Ni layers.\cite{mishra_2021}  In the work reported here, $d_{F1}$ was fixed at 2nm for all samples.  The Nb(20)/Cu(2) base layer was chosen to simulate the base layer of the Josephson junctions. Magnetic measurements were performed at $T=10$K using a SQUID-VSM with the applied field oriented in the plane of the samples.

\begin{figure}[!htbp]
\includegraphics[width=\linewidth]{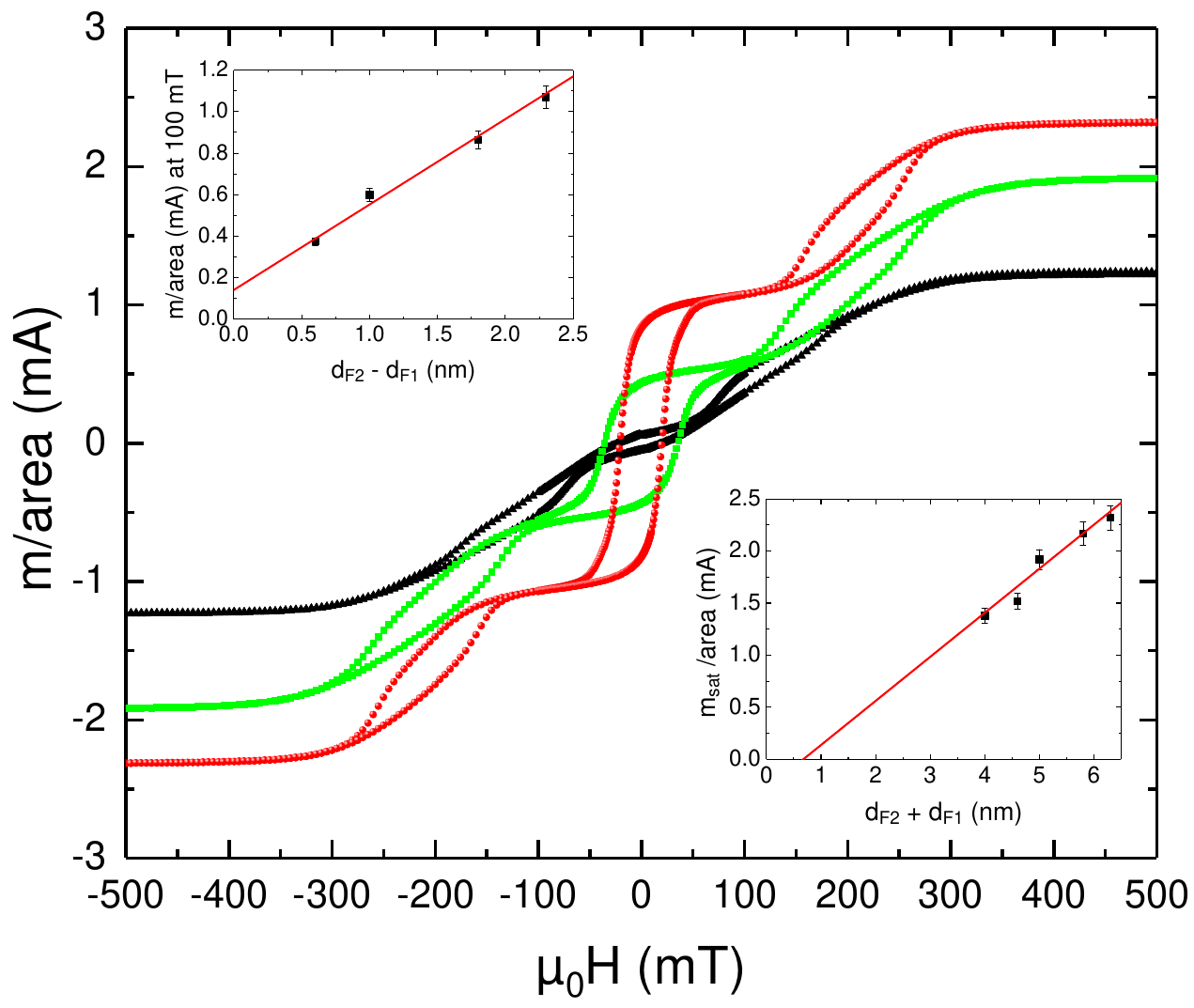}
\caption{Magnetic moment per unit area vs applied field for three representative Ni($d_{F1}$)/Ru(0.9)/Ni($d_{F2}$) SAF samples with $d_{F1}$ fixed at $2$nm and three values of $d_{F2}$: $2.0$nm (black points), $3.0$nm (green points), and $4.3$nm (red points). Left inset: Magnetic moment per area measured at $\mu_0H=100$mT vs $d_{F2}-d_{F1}$ for all the unbalanced SAF samples, along with a linear fit to the data.  Right inset: Magnetic moment per area measured at saturation ($\mu_0H=600$mT) vs $d_{F1}+d_{F2}$ for all the SAF samples, along with a linear fit to the data.} 
\label{fig:MvH}
\centering
\end{figure}

Figure \ref{fig:MvH} shows data of magnetic moment per unit area vs applied field $H$ for three samples with $d_{F2}=2.0$, 3.0, and 4.3 nm.  We also fabricated and measured samples with $d_{F2}=$2.6 and 3.8 nm, but omitted them from the main figure for clarity.  The samples containing unbalanced SAFs (i.e. all except for the sample with $d_{F2}=2.0$ nm) exhibit two salient features, one in the low-field range $|\mu_0H|<100$ mT and another in the intermediate-field range $100\mathrm{mT}<|\mu_0H|<300$ mT.  The low-field feature indicates where the magnetization of the entire SAF switches its direction so that the magnetizations of the thicker and thinner Ni layers point parallel and antiparallel to $H$, respectively.  That interpretation is confirmed by the plot in the left inset, which shows the moment per unit area measured at the field $\mu_0H=100$ mT vs $d_{F2}-d_{F1}$.  Although there is some scatter in the data, the straight line fit passes close to the origin, confirming that the net magnetization of the SAFs at that field is nearly proportional to the difference between the moments of the two Ni layers due to their antiparallel coupling. The slope of the fit yields an estimate of the Ni magnetization of $411\pm34$ emu/cm$^3$.  The somewhat large coercivity of the SAFs is due to the magnetocrystalline anisotropy of the individual Ni layers as well as extrinsic factors such as pinning at defects and surfaces.  Nevertheless, the coercivity of these unbalanced SAFs is noticeably less than that of individual Ni films with thicknesses $d_F=d_{F2}-d_{F1}$.\cite{klose_2012,baek_2014}  

The feature at intermediate field in Fig. \ref{fig:MvH} indicates where the two Ni magnetizations are being forced to align parallel to each other as the Zeeman energy overcomes the antiparallel exchange coupling.  That interpretation is confirmed by the plot in the right inset, which shows moment per area measured at saturation ($\mu_0H=600$ mT) vs $d_{F1}+d_{F2}$.  The straight line fit passes close to the origin, confirming that the saturated moment per area is proportional to the total Ni thickness in the samples.  The slope of this fit yields a second estimate of the Ni magnetization equal to $423\pm41$ emu/cm$^3$. The large uncertainty in the horizontal intercept of $0.7\pm0.5$nm precludes us from drawing any conclusions about the magnetic dead layer thicknesses at the Ni/Ru and Ni/Cu interfaces.  While the magnetization values extracted from the data shown in the two insets are lower than that of bulk Ni, they are consistent with reduced magnetizations observed in S/F systems by other groups.\cite{obi_1999,mattson_1997}  

Josephson junction samples were fabricated using a process similar to that in our previous works.\cite{niedzielski_2018}  We start by sputtering multilayers of the form [Nb(25)/Au(2.4)]$_3$/Nb(20)/Cu(2)/Ni($d_{F1}$)/Ru(0.9)/
Ni($d_{F2}$)/Cu(2)/Nb(5)/Au(5) on $1.27\times1.27$cm$^2$ Si chips coated with photolithography patterns defining the sample area and all leads and contact pads.  Then, six elliptically-shaped Josephson junctions with dimensions 1.25$\mu$m$\times$0.5$\mu$m are defined on each chip by e-beam lithography and ion milling, using two layers of ma-N2401 negative e-beam resist.  (The two layers were found to aid in the subsequent lift-off procedure.)  After ion milling the junctions are coated with 45 nm of evaporated SiO$_x$ for electrical insulation of the bottom leads, and then the ebeam resist is removed.  Finally, the top Nb(150)/Au(5) electrode is sputtered through another photolithography pattern after a gentle \textit{in-situ} ion mill.  

Transport measurements on the Josephson junctions are performed using a home-built dip-stick in a liquid helium storage dewar equipped with a Cryoperm magnetic shield. A magnetic field is applied in the plane of the junctions along the long axes of the elliptical junctions.  The field is swept between $\pm80$mT in steps of 2 mT.  $I-V$ curves are measured at each field value and fit with the standard form for overdamped Josephson junctions, the Resistively Shunted Junction model:\cite{barone_1982}
\begin{equation}
    V = \mathrm{sign}(I) R_N \Re\left\{\sqrt{I^2-I_c^2}\right\}
\end{equation}
where $I_c$ is the critical current, $R_N$ is the normal-state resistance of the junction and $\Re$ represents the real part of the argument. Values of $I_c$ and $R_N$ are obtained from fitting the above equation to the experimental data at each value of the applied magnetic field.

The magnetic field dependence of $I_c$ is expected to follow an Airy function \cite{barone_1982} for elliptically shaped junctions when the field is applied along a principal axis:
\begin{equation} \label{Eqn:Airy}
    I_c(\Phi) = I_{c0} \left| \frac{2 J_1 \left( \frac{\pi \Phi}{\Phi_0} \right) }{\frac{\pi \Phi}{\Phi_0}} \right|
\end{equation}
where $I_{c0}$ is the maximum value of $I_c$, $J_1$ is the Bessel function of the first kind, and $\Phi_0 = 2.07 \times 10^{-15}$ Tm$^2$ is the flux quantum.  The total magnetic flux through the junction includes a contribution from the net magnetization of the Ni/Ru/Ni SAF, and is given approximately by:
\begin{equation} \label{Eqn:Flux}
    \Phi = \mu_0 H w (2\lambda_{\mathrm{eff}} + d_N + d_{F1} + d_{F2}) + \mu_0 M w (d_{F2} - d_{F1})
\end{equation}
where $\lambda_{\mathrm{eff}}$ is the effective London penetration depth of the superconducting electrodes and $d_N$ is the thickness of the normal (non-ferromagnet/non-superconductor) layers.  Equation (\ref{Eqn:Flux}) should be valid in the low field range where two Ni magnetizations in the SAFs are coupled antiparallel to each other, as confirmed by the data in Fig. \ref{fig:MvH}. We thus expect the Airy patterns to be shifted in field by an amount
\begin{equation} \label{Eqn:Hshift}
    \mu_0 H_{\mathrm{shift}} = \frac{-\mu_0 M_{\mathrm{Ni}} (d_{F2}- d_{F1})}{(2\lambda_{\mathrm{eff}} + d_N + d_{F1} + d_{F2})}
\end{equation}

\begin{figure}[!htbp]
\includegraphics[width=\linewidth]{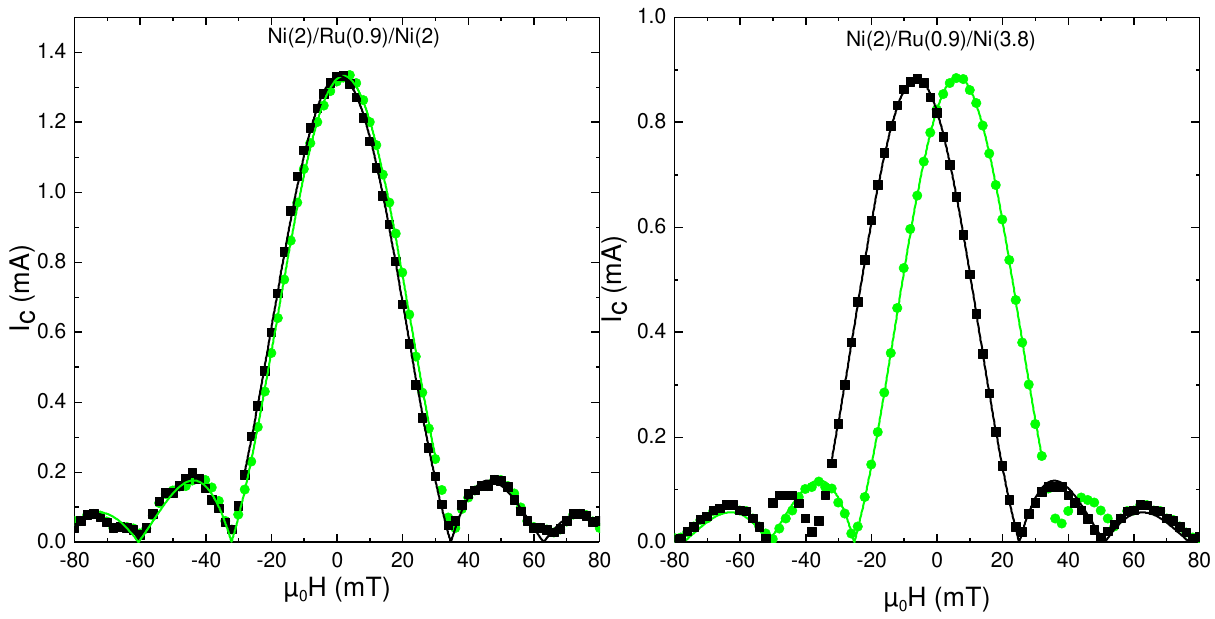}
\caption{Critical current ($I_c$) vs field ($H$) for Josephson junctions containing Ni($d_{F1}$)/Ru(0.9)/Ni($d_{F2}$) SAF samples with $d_{F1}$ fixed at $2$nm and $d_{F2}=2.0$nm (left) or $d_{F2}=3.8$nm (right). The black and green squares represent the data taken during downsweep and upsweep of the magnetic field, respectively. The black and green solid lines are fits to Eqn. \ref{Eqn:Airy} for the downsweep and upsweep data, respectively. The junction containing the balanced SAF (left) exhibits little magnetic hysteresis, while the junction containing the unbalanced SAF (right) exhibits marked hysteresis. These samples are representative of all the measured junctions.}
\label{fig:Fraunofers}
\centering
\end{figure}

Fig. \ref{fig:Fraunofers} shows the field dependence of $I_c$ for two junctions, one containing a balanced SAF and the other containing an unbalanced SAF. The black and green points represent data taken during downsweep and upsweep of the magnetic field, respectively, while the black and green solid lines are fits to the data of the Airy function, Eqn. (\ref{Eqn:Airy}). As expected, the data for the junction with the balanced SAF exhibit very little magnetic hysteresis, while the junction with the unbalanced SAF shows significant hysteresis.  Starting from a field of +80 mT, the net magnetic moment of the unbalanced SAF points in the positive direction, hence the central peak of the Airy pattern is shifted to negative field. The data follow the Airy function very well until a field of about -50mT, where the magnetization of the SAF suddenly reverses direction. The upsweep data are nearly a mirror image of the downsweep data.  

\begin{figure}[!htbp]
\includegraphics[width=\linewidth]{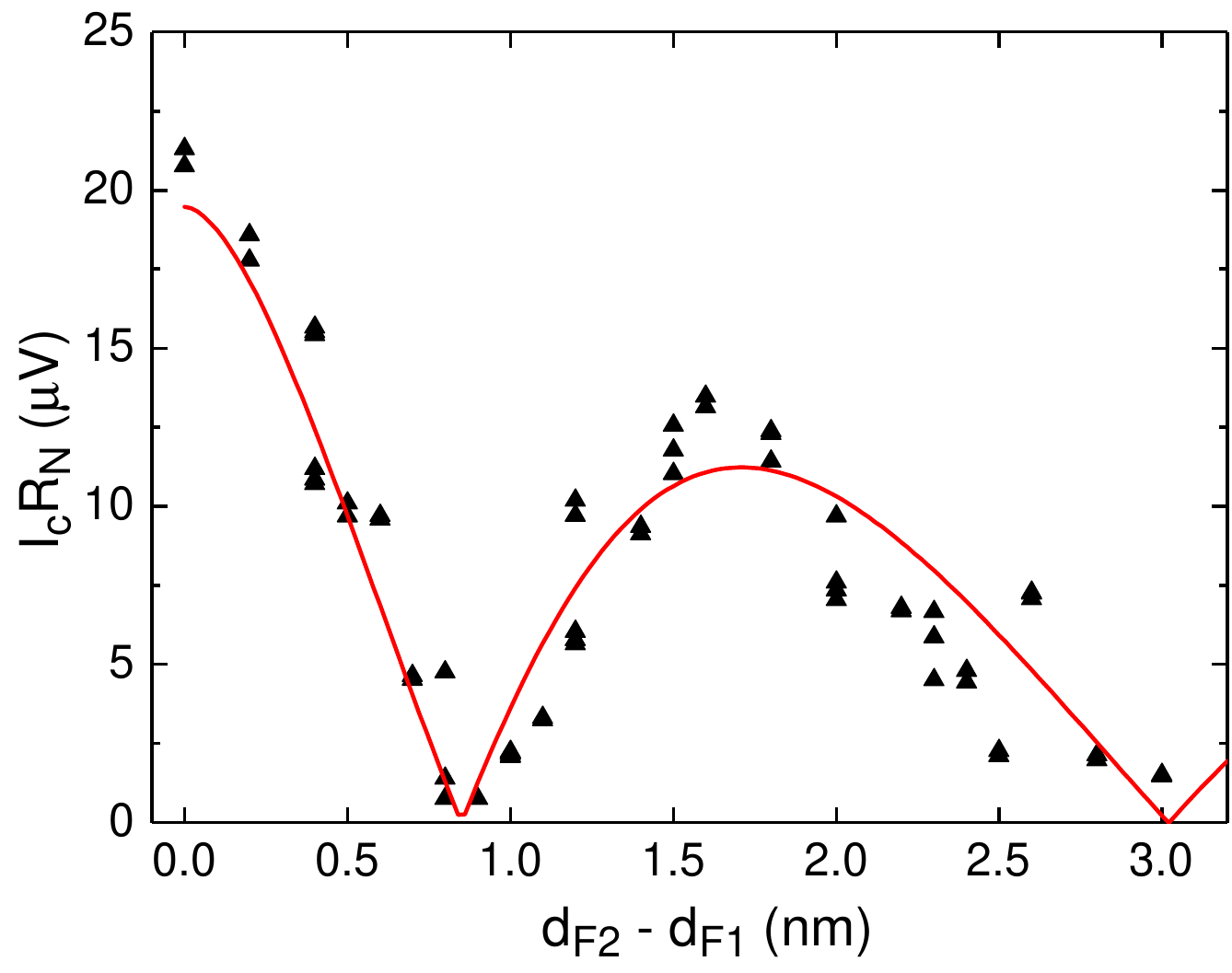}
\caption{$I_c R_N$ vs $d_{F2}-d_{F1}$ for all the measured Josephson junctions.  The solid line is a fit to the clean-limit theory discussed in the text.}
\label{fig:IcRn}
\centering
\end{figure}

The maximum values of $I_c$ extracted from the Airy function fits are multiplied by $R_N$ to get the $I_cR_N$ product for each junction measured.  $I_cR_N$ is a useful quantity because it is independent of junction area and facilitates comparison of junctions containing different materials or made by different research groups. Figure \ref{fig:IcRn} shows a plot of $I_cR_N$ vs $d_{F2}-d_{F1}$ for all the junctions measured.  The data exhibit a pronounced minimum at $d_{F2}-d_{F1}=0.86$ nm, indicating a transition from standard junctions to $\pi$-junctions.  There is a hint of a second transition near the edge of the data range at $d_{F2}-d_{F1}\approx 3.0$ nm, but we would need junctions with larger values of $d_{F2}-d_{F1}$ to verify that.  To test the theoretical expectations discussed in the introduction, we compare our data with data from Josephson junctions containing only a single Ni layer of varying thickness.  While several groups have fabricated and measured Ni junctions,\cite{blum_2002,shelukhin_2006,robinson_2006,bannykh_2009,baek_2017,baek_2018,dayton_2018,tolpygo_2019,kapran_2021} the most comprehensive data set in the thickness range containing the first $0-\pi$ and $\pi-0$ transitions is that of Baek \textit{et al.}\cite{baek_2017,baek_2018}  The data shown in Fig. \ref{fig:IcRn} look remarkably similar to the data of Baek \textit{et al.} -- most notably the location of the first $0-\pi$ transition at a thickness difference $d_{F2}-d_{F1} \approx 0.9$ nm.  That observation is the main result of this work, and confirms the theoretical expectation that the Ni/Ru/Ni SAF acts nearly equivalently to a single Ni layer with thickness $d_{F2}-d_{F1}$.\cite{blanter_2004}

The main difference between our data and the data of Baek \textit{et al.} is that our critical currents are significantly smaller. For example, 
our values of $I_cR_N$ near the peak of the $\pi$ state are in the range of $10-13\mu$V, whereas those of Baek \textit{et al.} are nearly an order of magnitude larger.  That is undoubtedly due to the presence of the Ru spacer in our samples, which adds two Ni/Ru interfaces to the total structure. 

Baek \textit{et al.} fit their data with the clean-limit theory of Buzdin, Bulaevskiy and Panyukov,\cite{buzdin_1982}, and find excellent agreement with the data. We have done the same; the solid curve in Fig.  \ref{fig:IcRn} is a fit of Eqn. (3) in ref. [\citenum{buzdin_1982}] to the data, using as free parameters only the overall amplitude of $I_cR_N$ and the ferromagnetic coherence length $\xi_F=(\delta k)^{-1}\approx\hbar v_F/2E_{Ex}$, where $\delta k=k_{F\uparrow}-k_{F\downarrow}$ is the wavevector shift between majority and minority spin electrons at the Fermi level. We obtain the value $\xi_F=0.81$ nm, which agrees with the wavevector shift obtained from photoemission data on Ni: $\delta k=1.2 \pm0.1$ nm$^{-1}$.\cite{petrovykh_1998}  It is common to add a third free parameter to such fits to account for magnetic dead layers; that was not needed to produce the curve shown in the figure.  That result is perhaps not surprising, because each of the Ni layers in our junctions are surrounded by Ru on one side and Cu on the other. Since the overall Cooper pair phase shift is determined by the difference between the phase shifts in the two Ni layers that make up the SAF, any magnetic dead layers should exactly cancel. 

More surprising, from our point of view is the precise coincidence of the position of our first $0-\pi$ transition with that of Baek \textit{et al.}\cite{baek_2017,baek_2018} It is easy to think of reasons why the location of the $0-\pi$ transition might be shifted in our junctions relative to junctions containing a single Ni layer. Theoretically, it is known that introducing extra normal metal spacer layers or changing the transmission at interfaces can shift the position of the $0-\pi$ transition significantly.\cite{pugach_2011,heim_2015} Clearly the Ni/Ru interfaces suppress supercurrent, probably due to Fermi surface mismatch; why don't they shift the transition?  Baek \textit{et al.} also measured Ni junctions that contained an additional nonmagnetic disordered alloy -- (Ni$_{81}$Fe$_{19}$)$_{69}$Nb$_{31}$, and found that the position of the first $0-\pi$ transition shifted from a Ni thickness of 0.9 nm to 1.5 nm. One might have thought that we would observe a similar shift. 

We note that our data can also be fit with a different model intended for junctions in the ``moderately clean limit," as described in the S.I. of ref. [\onlinecite{robinson_2010}].\cite{buzdin_private} That model predicts an oscillation and exponential decay of $I_cR_N$, with a decay length given by the mean free path. Over the range of data shown in Fig. 3, the two models are indistinguishable. The price to pay with the second model is the addition of two more free parameters: the mean free path and a parameter to account for the position of the first $0-\pi$ transition. The value of $\xi_F$ obtained from the fit using the second model is 0.68 nm -- close to the value given from fitting the first model.  But the value of the mean free path obtained from the fit is only $2.6$ nm, which is significantly shorter than the value of 7.5 nm obtained in an earlier study of Josephson junctions containing balanced Ni/Ru/Ni SAFs.\cite{mishra_2021} Given the ambiguity of the fitting procedure, we refrain from claiming ballistic transport in our junctions, and merely emphasize the striking similarity between our data and data taken from junctions containing a single Ni layer.

Regarding technological applications, these junctions have one important drawback, namely the substantial reduction in critical current. It might pay to look for a different coupling material than Ru,\cite{parkin_1991} with the hope of finding one with adequately large coupling that suppresses the supercurrent less. 

\section*{AUTHOR DECLARATIONS}

\subsection*{Conflict of Interest}
The authors have no conflicts to disclose.

\subsection*{Author Contributions}

\noindent\textbf{D. Korucu}: Investigation (lead); Data curation (equal); Formal analysis (equal); Writing (supporting). \textbf{R. Loloee}: Investigation (supporting); Data curation (equal); Formal analysis (equal). \textbf{N.O. Birge}: Conceptualization (lead); Data curation (supporting); Formal analysis (supporting); Writing (lead).

\section*{Data Availability}

The data that support the findings of this study are available from the corresponding author upon reasonable request.

\begin{acknowledgments}
D.K. thanks the AAUW for financial support. We thank A.I. Buzdin, R. Klaes, and N. Satchell for helpful discussions, D. Edmunds and B. Bi for technical assistance, and we acknowledge the use of the W. M. Keck Microfabrication Facility at Michigan State University. 
\end{acknowledgments}

\bibliography{Korucu_refs}

\begin{thebibliography}{49}
\bibitem [{buz()}]{buzdin_private}%
  \BibitemOpen
  \href@noop {} {}\bibinfo {note} {Ref. [18] provides formulas valid only for the special case $d_{F1}=d_{F2}$, but the model can be generalized to the case $d_{F1}\neq d_{F2}$. The result is that the oscillating term contains ($d_{F2}-d_{F1}$) while the exponentially decaying term contains ($d_{F1}+d_{F2}$). A.I. Buzdin, private communication}\BibitemShut {NoStop}%
\end{thebibliography}

\end{document}